# Title: Generative AI as a Learning Buddy and a Teaching Assistant: Preservice Teachers' Use and Attitudes

Matthew Nyaaba[1,2,3]*; Lehong Shi[1,2,4]; Macharious Nabang[5] Patrick Kyeremeh[6]; Xiaoming Zhai[1,2,7]; Samuel Arthur Ayoberd[8]; Bismark Nyaaba Akanzire[9]

[1] AI4STEM Education Center, University of Georgia, Athens, GA, USA
[2] Generative AI for STEM+C (GeNIUS+ C) Project, University of Georgia, Athens, GA, USA
[3] Department of Educational Theory and Practice, University of Georgia, USA
[4] Department of Workforce Education and Instructional Technology, University of Georgia, USA
[5] Department of Creative Arts, Bagabaga College of Education, Tamale, Ghana
[6] Department of Mathematics & I.C.T., St. Joseph's College of Education, Bechem, Ghana
[7] Department of Mathematics, Science, and Social Studies Education, University of Georgia, USA
[8] Department of Science Education, University for Development Studies, Tamale, Ghana
[9] Department of Education, Gambaga College of Education, Gambaga, Ghana

**Abstract**

This cross-sectional study investigates how preservice teachers in the Global South interact with Generative Artificial Intelligence (GenAI) across academic and teaching tasks, despite facing infrastructural constraints such as limited internet access and high data costs. The study involved 167 preservice teachers across four teacher education institutions in Ghana. Descriptive and inferential analyses, including multiple and ordinal logistic regressions, reveal that preservice teachers use GenAI as their learning buddy for tasks such as finding reading materials, seeking in-depth content explanations, and finding practical examples. Furthermore, they use GenAI as a teaching assistant for instructional tasks, such as finding teaching resources, identifying assessment strategies, and developing lesson objectives. Their usage patterns demonstrate that preservice teachers at the advanced levels of their programs (3rd year and 4th year) have significantly higher use frequencies of GenAI tools across these tasks. Interestingly, gender did not emerge as a significant predictor of their GenAI use, unlike their class level and age. They exhibited positive attitudes toward GenAI, indicating that it supports them in being more autonomous learners without relying heavily on peers and teachers for routine academic and teaching preparation tasks. However, they identified, among others, the cost of internet data, the inaccuracy of GenAI outputs, and concerns about academic dishonesty as significant challenges limiting their frequency of use. We therefore recommend GenAI literacy, focusing on ethical and responsible AI integration in teacher education in the Global South.

## Introduction

Recent research highlights the rapid rise of Generative Artificial Intelligence (GenAI) in education, with teacher education in the Global South also beginning to experience its growing influence, given the region's unique challenges and opportunities (Mbambo & Du Plessis, 2024). The Global South has unique contexts and faces numerous challenges in engaging with AI technology, particularly when operating with limited resources to inform global research conversations. Recent studies show that teacher educators in the Global South are generally positive about using GenAI in teacher education (Akanzire et al., 2025). However, they are concerned about how preservice teachers currently use these tools, particularly in relation to ethical issues and detecting whether preservice teachers rely on GenAI for assessments (Mbambo & Du Plessis, 2024; Nyaaba & Zhai, 2024b). Not much is currently known about preservice teachers' usage patterns of GenAI, especially in the Global South. A few emerging studies have begun to document how preservice teachers in the Global South are turning to GenAI tools to support academic tasks such as action research, an area often considered especially challenging (Nyaaba, Kyeremeh, et al., 2024). In Ghana, which serves as the contextual focus of this study, preservice teachers enrolled in Bachelor of Education (B.Ed.) programs engage in both academic coursework and structured teaching practice from their first year through to graduation.

A systematic review conducted by Salas-Pilco et al. (2022) on AI in teacher education emphasized the need for more research focused on the attitudes of both pre- and in-service teachers toward the use of GenAI in their learning and teaching practices. Following this, the present study investigates the specific learning and instructional tasks for which preservice teachers employ GenAI in teacher education in the context of Ghana, as well as the use pattern across academic learning and instructional preparation. For clarity, we conceptualize GenAI's use in academic learning as functioning as a “learning buddy”, supporting tasks such as explaining concepts, accessing personalized content, and completing coursework. Similarly, GenAI’s use in instructional preparation is framed as functioning as a “teaching assistant”, assisting with lesson planning, identifying assessment strategies, and generating teaching resources (Ibarra-Sáiz et al., 2025; Kiyak & Kononowicz, 2024; Lee & Zhai, 2024).

Additionally, recent studies have shown that background factors such as GenAI anxiety, social influence, age, year of study, gender, and performance expectancy significantly predict preservice teachers’ attitudes toward using GenAI (Wang et al., 2024). To understand the attitudes of Global South preservice teachers toward using GenAI, we further explore their attitudes toward integrating GenAI into teacher education as a learning buddy and teaching assistant, and examine potential demographic factors, such as age, gender, and year of study, that may predict their attitudes and patterns of GenAI use. Given that flagship journals such as the *Journal of Teacher Education*, in preparation for its Diamond Anniversary in 2025, have called for “bold new visions for the future of teacher education” and urged the field to “embrace the digital age and the integration of AI” to transform pedagogical landscapes, we believe the findings of this study will contribute meaningfully to that ongoing conversation (Wilder et al., 2024, p. 476). Specifically, it sheds broader light on how preservice teachers are engaging with GenAI and helps establish a foundation for future research in this area. Additionally, the study has the potential to inform global policy discussions across regions.

***Research Questions***

1. What do preservice teachers use GenAI for in specific learning (as a learning buddy) and instructional tasks (as a teaching assistant) in teacher education?

2. How frequently do preservice teachers engage with GenAI in their academic learning and instructional preparation?
3. What are preservice teachers' attitudes toward the integration of GenAI in teacher education?
4. What demographic factors predict preservice teachers' attitudes toward and the frequency of GenAI use in teacher education programs?

## Literature Review

### *Preservice Teachers' Attitudes and Experiences with Generative AI*

Recent research highlights that while preservice teachers are generally optimistic about the potential of GenAI, including ChatGPT, in lesson planning and teaching, their experiences reveal critical shortcomings that temper this enthusiasm. For instance, Slovak EFL preservice teachers expressed optimism about AI's role in English language learning despite limited understanding of its computational principles (Pokrivcakova, 2023), and Chinese preservice teachers reported that GenAI tools like ChatGPT enhanced classroom management and inclusive learning through personalized instruction (Zhang & Zhang, 2024; Nyaaba et al., 2024). At the same time, empirical work has documented persistent challenges when preservice teachers critically engage with AI outputs. Collaborative lesson design research indicates that ChatGPT can meaningfully enhance pedagogy and content when students possess sufficient prior knowledge and engage in higher-level collaboration (Okulu & Muslu, 2024; Wang, Zheng, & Chen, 2024). Similarly, Kalenda et al. (2025) found that 59 preservice teachers in STEM, TESOL, and Social Studies initially believed ChatGPT could generate complete and detailed lesson plans. Still, their perceptions declined after analysis revealed omissions in cultural responsiveness, scaffolding, differentiation, and procedural clarity. These findings align with results from mathematics and science contexts, where preservice teachers noted that ChatGPT outputs were often misaligned with curriculum standards and inadequate for fostering student engagement (Şimşek, 2025; Dimas et al., 2025).

Beyond lesson planning, many preservice teachers have already been using GenAI, sometimes without formal approval for research-related tasks, particularly in designing literature reviews and writing action research papers, which they perceive as challenging requirements of their programs (Aaron, 2018; Nyaaba et al., 2024). While this underscores GenAI's perceived utility, it also explains the concerns of teacher educators about academic dishonesty and the erosion of research competence (Nyaaba & Zhai, 2024b; Aaron, 2018). Broader adoption studies further emphasize that sustained use depends not only on individual perceptions but also on institutional support and social influence (Wang & Xie, 2025). These findings illustrate both the promise and tensions of GenAI integration: while preservice teachers remain optimistic about its potential to enhance learning and teaching, responsible adoption requires structured training, critical evaluation, and supportive policies that position AI as a complement rather than a replacement in teacher education.

### *Technological Pedagogical Content Knowledge and AI Integration*

The Technological Pedagogical Content Knowledge (TPACK) framework, originally conceptualized by Mishra and Koehler (2006), provides a comprehensive model for understanding how teachers integrate technology into pedagogy and subject matter instruction. Rooted in the interplay between Technological Knowledge (TK), Pedagogical Knowledge (PK), and Content Knowledge (CK), TPACK has evolved to adapt to subject-specific domains and emerging educational technologies. In response to the rising significance of GenAI in education,

Yue et al. (2024) introduced a revised TPACK model tailored explicitly for AI education. In their adaptation, TK demonstrates the ability to navigate and utilize AI tools and platforms effectively, including an awareness of their capabilities and limitations. CK encompasses the foundational principles of AI, such as algorithms, data literacy, machine learning, and ethics. PK, as in the original framework, pertains to general instructional knowledge, including strategies for effective classroom engagement and student learning.

Their framework also redefines the intersections of TPACK and AI:

- *TCK* highlights how AI tools can be leveraged to simulate or model content knowledge (e.g., using AI to visualize algorithms).
- *TPK* emphasizes how AI technologies can support teaching strategies, such as personalization or scaffolding.
- *PCK* addresses how educators can make AI content accessible and engaging.
- The unified TPACK domain encapsulates the integrated expertise needed to design and implement AI-rich learning environments.

This AI-specific TPACK model directly shapes the framework for our developed instrument for data collection and discussion in this study (Yue et al., 2024). It enabled us to classify how preservice teachers might use GenAI in their teacher education program, specifically, through two emerging themes: *GenAI as a learning buddy* and *GenAI as a teaching assistant*. The "learning buddy" function corresponds to uses aimed at enhancing personal understanding and inquiry (TK, PK, CK), such as clarifying complex theories, finding learning materials, or generating new concepts. The "teaching assistant" role captures GenAI's use in instructional planning, such as lesson design, generating assessment items, and drafting feedback, which align with Yue et al. (2024) TPK, TCK, and PCK domains.

### *Determinants of Generative AI Use*

Prior research has demonstrated that a complex interplay of individual, demographic, and structural factors shapes attitudes toward AI. Demographic predictors such as age, gender, academic year, and digital fluency are particularly salient: younger and more advanced students, as well as those with higher technological proficiency, show greater openness to GenAI (Ma & Lei, 2024; Sebetci, 2024), while male students often report higher confidence and fewer concerns (Al Omari et al., 2024). Race, language background, and identity also matter, with non-native English speakers and students from Asian and Black backgrounds reporting more academic benefits than their White peers (Deng et al., 2025). Studies show that personality traits, AI knowledge, and anxiety significantly influence positive and negative dispositions toward AI (Kaya et al., 2022), while preservice teachers' actual engagement with AI and their competence in AI-supported pedagogy predict their confidence and intention to integrate AI in classrooms (Harakchiyska, 2025; Liu et al., 2025). Although perceived usefulness, ease of use, and self-efficacy consistently predict adoption, concerns about trustworthiness and limited expertise remain barriers (Gamlem et al., 2025). At the same time, systemic barriers, including high data costs, unreliable internet connectivity, and limited institutional support, continue to hinder equitable adoption, particularly in under-resourced educational settings (Du & Gao, 2022; Ma & Lei, 2024). While the present study does not comprehensively address all these variables, it draws on insights from broader studies and focuses specifically on age, gender, and academic year as key determinants of GenAI use among preservice teachers.

## Method

We employed a quantitative cross-sectional survey design, a non-experimental approach that uses structured questionnaires to collect data at a single point in time (Setia, 2016). We selected this design because it is well-suited for addressing our research questions and our target participants, who are across different academic levels and institutions. Given the exploratory nature of this study and its rapid approach, this design was an efficient and cost-effective means of gathering large-scale, self-reported data (Setia, 2016).

### *Participants*

The participants in this study are preservice teachers enrolled in teacher education programs at four institutions in Ghana. A total of 167 preservice teachers participated in the study. Table 1 summarizes the demographic characteristics of the participants. The sample included most male respondents (73.1%), with most participants aged between 21 and 25 years (61.7%). In terms of academic level, more than half were in Level 400 [4$^{th}$ year in the program] (57.5%), followed by Level 300 [3$^{rd}$ year in the program] (35.9%).

**Table 1**

*Demographic Characteristics of Preservice Teacher Participants (N = 167)*

| Variable | Category | *f* (%) |
|---|---|---|
| Gender | Female | 45 (26.9) |
| | Male | 122 (73.1) |
| Age | 16-20 years | 4 (2.4) |
| | 21-25 years | 103 (61.7) |
| | 26-30 years | 54 (32.3) |
| | Above 30 years | 6 (3.6) |
| Class Levels | L200 | 11 (6.6) |
| | L300 | 60 (35.9) |

| | |
|---|---|
| L400 | 96 (57.5) |

### *Instrument*

We adapted and developed a structured instrument based on the validation framework proposed by Tzafilkou et al. (2021), which includes literature grounding, expert review, item revision, and construct validation. The instrument consisted of four sections. The first section gathered demographic data, including age, gender, and level of study (Al Omari et al., 2024; Howard, 2016; Ma & Lei, 2024). The second section assessed the frequency of GenAI usage in both learning and teaching, using a 5-point Likert scale. The third section then identified specific purposes, which we conceptually aligned with the AI-enhanced TPACK framework. Learning tasks (e.g., *reading materials*, *content explanations*) were categorized as *Learning Buddy* uses and correspond to TCK, as preservice teachers engaged GenAI to deepen their understanding of the subject matter. Reflective and assessment-related practices (e.g., *conducting reflections*, *identifying assessment strategies*) were framed within TPK, highlighting GenAI's role in supporting pedagogical strategies such as scaffolding, personalization, and metacognition. Teaching design tasks (e.g., *lesson planning*, *sample lessons*) were linked to PCK, reflecting the integration of pedagogy and content knowledge to make subject matter teachable and engaging. Finally, integrative activities (e.g., *locating instructional resources*) were situated within the unified TPACK domain, which emphasizes the holistic expertise required to design and implement AI-rich learning environments (Koehler & Mishra, 2009; Mishra, 2019). The final section measured preservice teachers' attitudes using a 12-item GenAI Attitude Scale. This scale initially consisted of 16 items and was subsequently modified to 12 items following the validation process. We conducted an *Exploratory Factor Analysis* (EFA) that revealed three constructs (Howard, 2016): *GenAI for Learning (GenAIL)* with six items, *GenAI for Teaching* (GenAIT) with four items, and *Ethical and Accuracy Aspects (EAA)* with two items. Items loaded strongly (≥ .40), consistent with Howard's (2016) loading rules. Internal consistency was excellent across factors: GenAIL ($\alpha = 0.91$), GenAIT ($\alpha = 0.90$), and EAA ($\alpha = 0.83$), exceeding the reliability benchmark of $\alpha \geq 0.70$ (Cho & Kim, 2015).

### *Data Collection*

The data collection process began with an orientation session to introduce the purpose of the study, ensure informed consent, and guide participants through the survey structure. The survey was self-administered via Google Forms, and links were distributed through class WhatsApp groups and course instructors. The data collection occurred at a time when Level 100 had not been reported to the school, so it involved preservice teachers in their level 200 [2 year], 300 [3rd year], and 400 [4th year]. A maximum period of four weeks was allocated for data collection, allowing sufficient time for participation, including follow-up reminders and opportunities for make-up submissions. Participation was voluntary, and respondents were assured of anonymity and confidentiality throughout the data collection and analysis processes.

### *Analysis*

Data analysis was conducted using SPSS Version 23 to address the four research questions (RQs) of the study. For *RQ1* and *RQ2*, which examined how and how frequently preservice teachers use GenAI as a *learning buddy* and a *teaching assistant*, descriptive statistics (frequencies and percentages) summarized the distribution of use across defined tasks. Additionally, item-level means *(M)* and standard deviations *(SD)* were calculated to capture patterns in frequency of use across both domains. *RQ3*, focusing on attitudes toward GenAI, was addressed using item-wise descriptive analyses for constructs related to learning, teaching, and ethical-accuracy concerns. To answer *RQ4a*, which examined the extent to which demographic variables (age, gender, and class level) predict preservice teachers' attitudes toward GenAI, we conducted a multiple linear regression analysis. To address *RQ4b*, which investigated whether these same demographic factors predict the frequency of GenAI use in both learning and teaching contexts, we performed ordinal logistic regression analyses for GenAI as a learning buddy and as a teaching assistant, respectively. All statistical assumptions, including multicollinearity, normality, and proportional odds, were verified and met before analysis. The assumptions of normality, linearity, outliers, and multicollinearity were first assessed. Nonetheless, no serious violations of these assumptions were identified. For instance, the collinearity statistics revealed Tolerance range values (.810 – .901), which were not below .10, suggesting that the multicollinearity assumption has not been violated.

## Results

### ***How Preservice Teachers Use GenAI for Specific Learning and Instructional Tasks***

In response to RQ1, the results reveal distinct usage patterns for both learning and instructional purposes, highlighting GenAI's role in enhancing preservice teachers' academic and teaching autonomy, as well as their access to supplementary content.

#### *Use GenAI as a Learning Buddy*

Regarding the participants who use GenAI as their *Learning buddy* (See Table 2), they primarily used GenAI to find reading materials (38.11%, $n$ = 93), followed by seeking in-depth content explanations (25.41%, $n$ = 62). Participants also used GenAI to find practical examples (15.16%, *n* = 37), conduct reflections (11.07%, *n* = 27), and extend their perspectives or knowledge (10.25%, *n* = 25).

**Table 2**

*Learning Areas Where Preservice Teachers Use GenAI as a Learning Buddy*

| Learning areas | # of participants | % |
|---|---|---|
| Extending Perspectives/Knowledge (TCK) | 25 | 10.25 |
| Seeking In-depth Content Explanations (TCK) | 62 | 25.41 |
| Finding Practical Examples (TCK) | 37 | 15.16 |
| Finding Reading Materials (TCK) | 93 | 38.11 |
| Conducting Reflections (Unified TPACK) | 27 | 11.07 |

*Use GenAI as a Teaching Assistant*

In terms of using GenAI as a *Teaching Assistant* (See Table 3), the participants indicated using GenAI mostly to find teaching resources (33.33%, *n* = 90). Other prevalent uses included identifying assessment strategies (21.48%, *n* = 58), identifying lesson objectives (16.30%, *n* = 44), creating lesson plans (15.93%, *n* = 43), and locating sample lessons (12.96%, *n* = 35). This pattern reflects GenAI's integration into preservice teachers' instructional planning and professional learning processes.

**Table 3**

*Instructional Areas Where Preservice Teachers Use GenAI as a Teaching Assistant*

| Instructional area | # of participants | % |
|---|---|---|
| Identifying Assessment Strategies (TPK) | 58 | 21.48 |
| Identifying Lesson Objectives (TPK) | 44 | 16.30 |
| Creating Lesson Plans (PCK) | 43 | 15.93 |
| Finding Sample Lessons (TPK) | 35 | 12.96 |
| Finding Teaching Resources (Unified TPACK) | 90 | 33.33 |

***Frequency of Generative AI Use by Preservice Teachers***

RQ 2 explored the frequency with which preservice teachers employ GenAI across academic learning and instructional tasks. The descriptive results presented in Figure 1 show distinct patterns in usage across the two contexts. As a learning buddy, GenAI is used with greater frequency: 28.14% of preservice teachers reported using it “Very Often,” 16.17% “Often,” and 43.11% “Sometimes,” totaling over 87% using it at least occasionally. In contrast, the use of GenAI as a teaching assistant was less frequent. Only 14.97% reported using it “Very Often” and 18.56% “Often,” while 36.53% used it “Sometimes.” Notably, 19.16% indicated “Never” using it for instructional purposes, and 10.78% used it “Rarely.” These findings suggest that preservice teachers are more inclined to use GenAI as a tool for supporting their own academic learning than for instructional or pedagogical tasks. This may reflect varying levels of confidence, access, or training in applying GenAI within teaching contexts compared to self-directed learning environments.

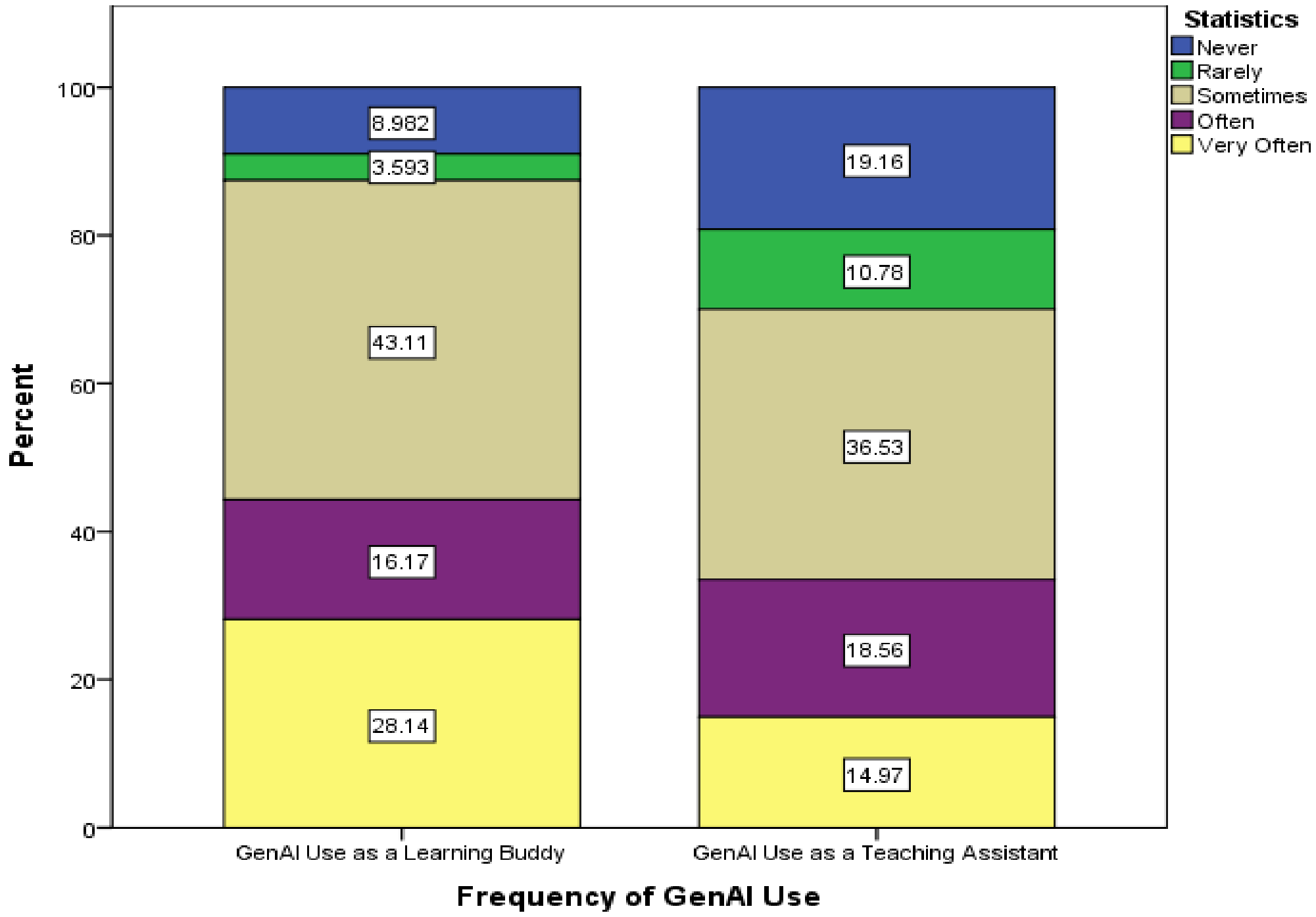

Figure 1: *Frequency of Generative AI use as Learning Buddy and Teaching Assistant*

### *Preservice Teachers' Attitudes Toward Generative AI*

RQ 3 sought to explore *preservice teachers' attitudes toward GenAI* across three dimensions: GenAIL, GenAIT, and EAA, as shown in Table 4. The results indicate generally positive attitudes across both the learning and teaching dimensions of GenAI use, with more nuanced perceptions regarding ethical and accuracy concerns. In the 'GenAIL' construct, the highest agreement was with the item *With the help of GenAI, I am able to get more reading materials* (M = 4.02, SD = 0.93), followed by *GenAI helped me understand various complex areas of my study* (M = 3.98, SD = 0.85). These suggest a strong endorsement of GenAI as a tool for enhancing access and comprehension. Conversely, the item *"Most students might cheat and will not develop critical and creative skills due to GenAI"* received a moderate mean score (M = 3.28, SD = 1.31), reflecting ambivalence about potential negative impacts on students' higher-order thinking. In the 'GenAIT' construct, the most endorsed item was "*GenAI has the potential to* *improve my pedagogical skills"* (M = 3.88, SD = 0.90), indicating perceived pedagogical value. The preservice teachers also showed moderate support for the integration of GenAI into teacher preparation programs (*M = 3.78, SD = 1.00*). However, under the 'EAA' construct, the lowest-rated item was "*the information provided by GenAI tools was not accurate enough, so I never trusted them"* (M = 2.62, SD = 1.07), suggesting a relatively higher trust in GenAI-generated content, though not without concerns. The item "*It is expensive using GenAI tools for teaching"* received a mid-level rating (M = 3.17, SD = 1.13), indicating cost might be a barrier for some participants.

**Table 4**

*Preservice Teachers' Attitudes Toward GenAI by Item-Level Descriptive Statistics*

| Construct | Item | *M* | *SD* |
|---|---|---|---|
| GenAIL | GenAI has the potential to positively improve students' content knowledge. | 3.86 | 1.08 |
| | GenAI helped me understand various complex areas of my study. | 3.98 | 0.85 |
| | With the help of GenAI, I am able to get more reading materials | 4.02 | 0.93 |
| | I don't require much assistance from colleagues for my studies since I am using GenAI | 3.35 | 1.17 |
| | GenAI explained things better to me than we had time to do in class | 3.49 | 1.13 |
| | Most students might cheat and will not develop critical and creative skills due to GenAI | 3.28 | 1.31 |
| GenAIT | GenAI has the potential to improve my pedagogical skills (teaching skills) of teachers | 3.88 | 0.90 |
| | I didn't/don't require much assistance from colleagues for my teaching since I am using GenAI | 3.44 | 1.16 |
| | I felt more confident preparing and teaching with GenAI's assistance. | 3.67 | 1.00 |
| | I suggest that GenAI tools be incorporated into our courses and teaching practice | 3.78 | 1.00 |
| EAA | It is expensive to use GenAI tools for teaching | 3.17 | 1.13 |
| | The information provided by GenAI tools was not accurate enough, so I never trusted them | 2.62 | 1.07 |

*Note.* Constructs: GenAIL = GenAI for Learning; GenAIT = GenAI for Teaching; EAA = Ethical and Accuracy Aspects.

### ***Prediction of Demographic Factors on GenAI Attitudes and Usage***

#### *Predicting Teachers' Attitudes towards GenAI*

A standard multiple regression was conducted to assess the ability of preservice teachers' age, gender, and class level to predict their attitudes towards GenAI. Table 5 shows that these demographic variables were not statistically significant in predicting preservice teachers' attitudes towards GenAI, $F(3, 163) = .319$, $p = .811$, $R^2 = .006$. These independent variables in the model explained only 0.6% of the variance in the dependent variable ($R^2 = .006$). The lack of statistical significance in the overall model ($F(3, 163) = 0.319$, $p = .811$) and the low $R^2$ value (0.006) indicate that the model does not adequately capture the relationship between the predictors and the outcome. This suggests that age, gender, and class level do not meaningfully predict preservice teachers' attitudes toward GenAI.

**Table 5**

*Multiple regression predicting preservice teachers' attitudes towards GenAI by Age, Gender, and Class Level*

| Predictor | *B* | *SE* | *95% CI* | *β* | *p* |
|---|---|---|---|---|---|
| Constant | 40.589 | 2.846 | [34.969, 46.209] | - | .000 |
| Age | .460 | .902 | [-1.322, 2.242] | .042 | .611 |
| Gender | .965 | 1.241 | [-1.485, 3.414] | .066 | .438 |
| Class Level | -.323 | .916 | [-2.132, 1.485] | -.031 | .725 |

*Note.* $R^2$ = .006, $F(3, 163)$ = .319, $p$ = .811[b]

*Predicting Frequency of GenAI Use*

*GenAI Use for Academic Learning (GenAIL).* The results show that younger preservice teachers and those at lower class levels reported less frequent use of GenAI for academic learning. In conducting the ordinal logistic regression analysis between the three independent variables (i.e., age, gender, and class level) and how often preservice teachers use GenAI in learning, the predictor variables were tested a priori to verify that there was no violation of the assumptions, including parallel lines and multicollinearity (See Table 6). The predictors accounted for a significant amount of variance in the outcome, as indicated by the likelihood ratio $\chi 2(6)$ = 16.651, p = .011. Specifically, the 16-20 years age group (B = -2.880, SE = 1.280, p = .024) and the Class Level 300 group (B = .886, SE = .342, p = .010) significantly predicted the frequency of preservice teachers' usage of GenAIL. Overall, the model accounted for approximately 4% of the variance in the outcome, with McFadden's pseudo-R-squared value of .037. Pearson's goodness-of-fit Test indicated a good fit, $\chi^2(58)$ = 51.645, p = .709.

**Table 6**

*Ordinal Logistic Regression Predicting How Often GenAI is used in Learning*

| Predictor (GenAIL) | *B* | *SE* | Wald $\chi^2$ | *p* | 95% *CI* |
|---|---|---|---|---|---|
| Threshold [GenAIL = 1] | -3.629 | .867 | 17.519 | .000 | [-5.329, -1.930] |
| Threshold [GenAIL = 2] | -3.237 | .855 | 14.324 | .000 | [-4.914, -1.561] |
| Threshold [GenAIL = 3] | -.926 | .825 | 1.258 | .262 | [-2.544, .692] |
| Threshold [GenAIL = 4] | -.167 | .822 | .041 | .839 | [-1.779, 1.444] |
| [Age (16-20years) =1] | -2.880 | 1.280 | 5.065 | .024 | [-5.388, -.372] |
| [Age (21-25 years) =2] | -1.355 | .842 | 2.588 | .108 | [-3.006, .296] |
| [Age (26-30 years) =3] | -1.333 | .856 | 2.426 | .119 | [-3.012, .345] |
| [Gender (Female) =1] | -.466 | .353 | 1.745 | .187 | [-1.157, .225] |
| [Class Level 200 =1] | -.724 | .640 | 1.279 | .258 | [-1.979, .531] |
| [Class Level 300 =2] | .886 | .342 | 6.707 | .010 | [.215, 1.556] |

*GenAI Use for Instructional Practice (GenAIT).* An ordinal logistic regression analysis was conducted to investigate the relationships between the three independent variables (i.e., age, gender, and class level) and the frequency with which preservice teachers use GenAIT. The predictor variables were tested a priori to verify that there was no violation of the assumptions, including parallel lines and multicollinearity. Together, the predictors accounted for a significant amount of variance in the outcome, as indicated by the likelihood ratio $\chi 2(6) = 16.905$, $p = .01$. According to Table 7, only Class Level 200, $B = -1.693$, $SE = .666$, p = .011, significantly predicted how often preservice teachers use GenAIT. Overall, the model accounted for approximately 3% of the variance in the outcome, with McFadden's pseudo-R-squared value of .033. Pearson's goodness-of-fit Test indicated a good fit, $\chi^2(58) = 48.571$, p = .807.

**Table 7**

*Ordinal Logistic Regression Predicting How Often GenAI is used in Teaching*

| Predictor (GenAIT) | *B* | *SE* | Wald $\chi^2$ | *p* | 95% *CI* |
|---|---|---|---|---|---|
| Threshold [GenAIT = 1] | -2.627 | .776 | 11.449 | .001 | [-4.148, -1.105] |
| Threshold [GenAIT = 2] | -1.987 | .766 | 6.726 | .010 | [-3.489, -.485] |
| Threshold [GenAIT = 3] | -.354 | .753 | .221 | .638 | [-1.829, 1.122] |
| Threshold [GenAIT = 4] | .737 | .755 | .952 | .329 | [-.743, 2.218] |
| [Age (16-20years) =1] | -1.789 | 1.194 | 2.244 | .134 | [-4.129, .551] |
| [Age (21-25 years) =2] | -1.051 | .768 | 1.872 | .171 | [-2.556, .454] |
| [Age (26-30 years) =3] | -1.025 | .785 | 1.707 | .191 | [-2.563, .513] |
| [Gender (Female) =1] | -.462 | .344 | 1.807 | .179 | [-1.135, .212] |
| [Class Level 200 =1] | -1.693 | .666 | 6.459 | .011 | [-2.999, -.388] |
| [Class Level 300 =2] | .425 | .328 | 1.682 | .195 | [-.217, 1.068] |

*Note*. *Pearson Chi-Square: $\chi^2(58) = 48.571, p = .807$)*

*Likelihood Ratio Chi-Square: $\chi^2(6) = 16.905, p = .01$*

*Test of Parallel Lines: $\chi^2(18) = 20.333, p = .314$*

## Discussion

The findings conceptualized preservice teachers' usage of GenAI as either a *learning buddy* (supporting personal learning) or a *teaching assistant* (aiding instructional planning) (Yue et al., 2024). The strong usage patterns as a learning buddy (e.g., accessing materials and explanations) suggest a solid grounding in AI-TCK (technological knowledge of AI tools). At the same time, its application in instructional planning (e.g., lesson planning and assessments) reflects emerging AI-TPK and AI-TCK. This aligns with the AI-TPACK framework's call for strengthening AI literacy (Yue et al., 2024). When used as a learning buddy, GenAI supported preservice teachers in locating relevant reading materials and providing detailed explanations of academic content. These practices suggest a high level of initiative and self-direction, which confirms prior findings by Lee and Zhai (2024) and Chen et al. (2024), who emphasized that preservice teachers leverage GenAI for autonomous learning and personalized academic support. Notably, in contexts such as the Global South, where access to traditional learning resources is often limited, GenAI has emerged as a meaningful alternative for supplementing gaps in textbooks, reference materials, and instructor feedback (Mbambo & Du Plessis, 2024).

Furthermore, the finding shows that preservice teachers in the Global South are currently using GenAI as their teaching assistant. They primarily use GenAI for identifying relevant teaching resources and assessment strategies, which align with similar findings by Mbambo and Du Plessis (2024), who reported that 115 teachers confirmed GenAI improved their instructional methods and resource accessibility. Additionally, this finding confirms the use of GenAI for instruction, as indicated by Zhang and Zhang (2024), which shows that 270 participants in teacher education programs acknowledged GenAI's capability to enhance classroom management and inclusive learning through personalized instruction. However, this study did not emphasize classroom management and inclusive education, as Zhang and Zhang (2024) did. In both cases, a notable observation is the variation in use between learning and teaching by the preservice teachers. While academic applications of GenAI were relatively frequent, like “sometimes” or “very often,” instructional uses were a little lower. Specifically, preservice teachers in higher-level classes (300 and 400) use GenAI more frequently, likely due to increased academic responsibilities, teaching practicum, or research projects.

In contrast, younger preservice teachers in this study or those earlier in their training showed less usage. This finding aligns with the scaffolding journey of teacher education programs, as not all preservice teachers engage in actual teaching activities until their third year of teaching (Ministry of Education, 2017). This finding has implications for how GenAI integration could follow a similar pattern to the established curriculum of most curricula, such as Ghana’s teacher education curriculum in this study. Moreover, the findings further support the assertion that preservice teachers in their advanced stages of their academic programs demonstrate greater openness and usage of GenAI tools for both learning and teaching purposes (Ma & Lei, 2024; Sebetci, 2024).

In terms of attitudes, the preservice teachers hold positive views of GenAI's integration in their teacher preparation program (Sarfo et al., 2024). They agreed strongly that GenAI can make them less dependent on their instructors and peers for support, yet expressed moderate concern over GenAI's potential to facilitate academic dishonesty or reduce the development of critical thinking skills. The findings align and confirm Sarfo et al.'s (2024) review study, which shows Sub-Saharan Africans, especially educators, generally have a positive perception towards GenAI; however, they have skepticism in GenAI integration due to financial constraints, just as the preservice teachers indicated that the cost of GenAI tools hinders their usage in this study. While attitudes toward GenAI did not vary significantly by age, gender, or class level, the frequency of use was influenced by these variables. Notably, no significant gender-based differences in usage were observed; however, this finding should be interpreted with caution, as the gender distribution in the sample was not balanced. Further studies with more equal representation are therefore needed to validate this observation. Nonetheless, the result aligns with previous research suggesting that male students often report higher confidence and express fewer ethical or pedagogical concerns regarding the adoption of GenAI (Al Omari et al., 2024). Furthermore, the findings suggest that preservice teachers are outpacing their instructors in GenAI usage, as many of them are independently exploring GenAI's potential more rigorously than their instructors, who are still calling for ethical approvals and development to start with in the Global South (Akanzire et al., 2025; Nyaaba & Zhai, 2024b). The study by Nyaaba and Zhai (2024b) reveals that many teacher educators lack GenAI literacies in prompt engineering and the effective use of GenAI in supporting their academic work, and therefore are calling for support in professional development.

The findings also indicated that the preservice teachers found the GenAI outputs to be inaccurate at times and struggled to trust them. This finding aligns with several studies on the accuracy and biases inherent in GenAI tools (O'Connor & Liu, 2024; Tao et al., 2024). This call for targeted and customized GenAI tools simultaneously requires critical AI literacy (Dilek et al., 2025; Giustini & Dastyar, 2024; Nyaaba & Zhai, 2024a). For instance, Nyaaba et al. (2024) emphasized the value of developing customized GenAI tools that streamline processes and provide targeted outputs.  Ultimately, these findings underscore the need to reimagine teacher education programs in ways that embrace and shape the future of GenAI integration in the Global South (Mbambo & Du Plessis, 2024). Rather than limiting GenAI access, teacher educators' programs should invest in building educators' capacity to guide students through responsible and effective GenAI use (Nyaaba & Zhai, 2024b). This includes fostering ethical decision-making, deepening critical thinking, and designing curriculum-aligned applications. As preservice teachers show increasing fluency in these tools, it becomes essential for teacher educators in the Global South to step up, not only to remain pedagogically relevant but to co-lead a new era of technology-enhanced education.

**Conclusion**

In this study, we examine how preservice teachers in the Global South, specifically those in Ghana, utilize GenAI in their teacher preparation programs. Preservice teachers in Ghana engage in four-year programs learning coursework and practicing teaching in partnering schools (known as supported teaching in schools). It is interesting to explore how these students, in an environment with limited resources and digital infrastructure, are navigating the use of GenAI in their teacher preparation. The findings show that preservice teachers use GenAI as a learning aid to assist them in academic tasks, such as finding reading materials, seeking content explanations, and finding practical examples. They also utilize GenAI as their teaching assistant, which assists them with lesson planning, assessment strategies, and sample lesson plans, among other tasks. There are high usage patterns among advanced class levels and older preservice teachers

Following these use patterns, the preservice teachers expressed positive attitudes toward GenAI integration in teacher education, given that GenAI assists them in being more autonomous in their academic learning and instructional tasks without heavily burdening peers and teachers. They, however, raise concerns about data costs and internet access, as well as the inaccuracy of GenAI output, as challenges to their effective use of GenAI in their tasks. These findings suggest that preservice teachers in the Global South have embraced and initiated the use of GenAI in teacher education. However, the unbalanced use pattern and attitudes among participants with different demographic background calls for the importance of promoting GenAI literacy among preservice teachers in the Global South to address and support their use. It calls for policies and guidelines to integrate GenAI use in teacher education programs in the Global South. The study also contributes to global use cases and lays the groundwork for further research in the Global South. We recommend future research to comprehensively explore the various use aspects and address the limitations of this study.

## Limitations

This study is subject to several limitations that should be taken into account when interpreting the findings. Firstly, the study relied on self-reporting methods, which can introduce biases such as social desirability and inaccurate recall. Participants may have over- or under-reported their usage of GenAI tools, especially in relation to their attitudes or frequency of use. Secondly, the study used a convenience sampling method, which introduces potential biases and limits the generalizability of the findings. The sample may not fully represent the broader population of preservice teachers in Ghana or other contexts in the Global South.

Additionally, there was an imbalance in the demographic distribution, particularly in gender representation, which could have influenced the responses and skewed interpretations in subtle ways. We further recommend that a more robust qualitative or mixed-methods approach could have provided richer insights into how preservice teachers engage with GenAI, particularly in under-resourced or highly contextualized teaching environments, such as those in the Global South. Future research should consider more in-depth qualitative approaches and balanced sampling to explore these emerging trends with greater depth and validity.